\documentclass[12pt]{iopart}
\usepackage{setstack}
\usepackage{amsbsy}
\usepackage{amstext}
\usepackage{iopams}
\usepackage{dsfont}
\usepackage{graphicx}

\renewcommand{\vec}{\boldsymbol}
\newcommand {\vecr}{\vec r}
\begin{document}
\title[Surface plasmon resonances of an arbitrarily shaped nanoparticle]{Surface
  plasmon resonances of an arbitrarily shaped nanoparticle: High frequency
  asymptotics via pseudo-differential operators}
\author{D Grieser$^1$ and F R\"uting$^2$}
\address{$^1$Institut f\"ur Mathematik, Carl von Ossietzky Universit\"at, 26111
  Oldenburg, Germany}
\ead{grieser@mathematik.uni-oldenburg.de}
\address{$^2$Institut f\"ur Physik, Carl von Ossietzky Universit\"at, 26111
  Oldenburg, Germany}
\ead{rueting@theorie.physik.uni-oldenburg.de}
\begin{abstract}
We study the surface plasmon modes of an arbitrarily shaped nanoparticle in the
electrostatic limit. We first deduce an eigenvalue equation for these modes,
expressed in terms of the Dirichlet-Neumann operators. We then use the
properties of these pseudo-differential operators for deriving the limit of the
high-order modes.
\end{abstract}
\maketitle

\section{Introduction}
The interaction between light and metallic nanoparticles can become very strong
due to the excitation of surface plasmons. These hybrid modes of the
electromagnetic field and the electron gas are confined to the surface of the
particle and give rise to an enhancement of the incident field by several orders
of magnitude~\cite{Kreibig_Vollmer,Bohren_Huffmann,Hao_Schatz}. This enhancement
enables a variety of applications ranging from the well-established
surface-enhanced Raman spectroscopy (SERS), which allows the detection of even a
single molecule \cite{Nie_Emory,Kneipp}, to the emerging field of plasmonics
\cite{Maier_Atwater,Ozbay}, including for instance plasmonic waveguides which
effectuate optical energy transfer below the diffraction
limit~\cite{Maier_Atwater,Brongersma,Maier}.\\

While an exact analytical description of the optical response of a metallic
nanoparticle exists only for very specific geometries like a sphere or an
ellipsoid, numerical methods can be applied for a particle with an arbitrary,
realistic shape. There is a wide range of numerical methods for light scattering
\cite{Wriedt}; the finite difference time domain approach (FDTD) is a very
common one (originally proposed by Yee~\cite{Yee}). FDTD combined with a
suitable discretisation of the particle enables one to calculate the response of
an almost arbitrary particle to nearly any incident field. Moreover, in the case
of a sufficiently small particle, i.e. for particles that are much smaller than
the significant wavelengths, for which the dielectric function can be taken as
constant, the surface plasmon resonances can be determined in the electrostatic
limit via the eigenvalues of a surface-integral operator
\cite{Abajo_Aizpurua,Abajo_Aizpurua2,Hohenester_Krenn}. Hohenester and Krenn
\cite{Hohenester_Krenn} have shown that this boundary integral approach can be
used to calculate the surface plasmon resonances of single and coupled spheres,
cylinders and cubic-like nanoparticles.\\

\begin{figure}[htbp]
\centering
\includegraphics[width=5 cm]{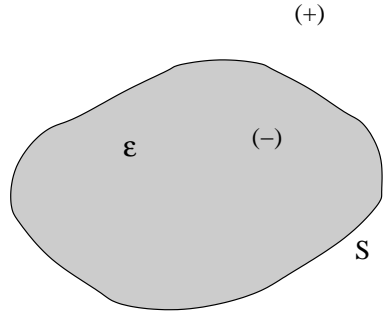}
\caption{Sketch of an example geometry, with quantities inside the particle being
  labelled by the index {\it -} and outside by {\it +}. {\it S} denotes the
  surface of the particle.}
\label{fig:geom}
\end{figure}

In this paper we consider an arbitrarily shaped nanoparticle (see
Fig.~\ref{fig:geom}) and show that the surface plasmon resonances in the
electrostatic limit may be obtained as eigenvalues of an operator which may be
represented in terms of Dirichlet-Neumann operators. We argue that this
reformulation makes the well-developed analytical tools pertaining to the study
of such operators available for the study of surface plasmon resonances in cases
where exact solution formulae are not available. To support this claim we show
here how the asymptotic behaviour of the high-order surface resonances may be
analysed. To this end we use the well-known fact that the Dirichlet-Neumann
operators are pseudo-differential operators, and use standard properties of such
operators.  \\

In Section~\ref{sec:eigenvalue} we reformulate the boundary value problem as an
eigenvalue problem and in Section~\ref{sec:half-space} we show how one can apply
the formalism to a half-space. In Section~\ref{sec:high} we introduce
pseudo-differential operators and prove in rigorous mathematical terms the
convergence of the high order modes.

\section{Eigenvalue equation}
\label{sec:eigenvalue}
In the electrostatic limit the surface plasmon resonances of a particle as
sketched in Fig.~\ref{fig:geom} are characterised by nontrivial solutions of
the Poisson equation without external charges,
\begin{eqnarray}
\Delta \phi_{\pm}(\vec{r})=0 &\text{ for } \vec{r} \not\in S, \label{eq:poisson}\\
\phi_{-}(\vec{r})=\phi_{+}(\vec{r}) &\text{ for } \vec{r} \in S, \label{eq:bc1}\\
\epsilon \partial_n \phi_{-}(\vec{r}) = \partial_n \phi_+(\vec{r}) &\text{ for } \vec{r} \in S, \label{eq:bc2}
\end{eqnarray}
with the potential $\phi_-$ and $\phi_+$ inside and outside the particle,
respectively; $\epsilon$ is the permittivity of the particle and $\partial_n$ is
the outward normal derivative on the surface $S$ of the particle. For
convenience we assumed by stating the boundary condition in eq.~(\ref{eq:bc2})
that the particle surrounded by vacuum is homogeneous, isotropic, local and
non-magnetic. Since we restrict ourselves to surface modes we further assume
that the potential vanishes for $r \to \infty$, so that $\phi$ is determined
uniquely. We assume that the surface $S$ is smooth, i.e.\ has no
singularities. So surface resonances are given by thus values of $\epsilon$ for
which nontrival solutions of the system (\ref{eq:poisson})-(\ref{eq:bc2}) exist.  \\

 In order to recast the problem (\ref{eq:poisson})-(\ref{eq:bc2}) as an
eigenvalue problem we introduce the Dirichlet-Neumann operators ($D_-$ and
$D_+$) inside and outside the particle. The Dirichlet-Neumann operator inside
the particle is defined as
\begin{eqnarray}
D_-:\quad C^{\infty}(S) \rightarrow C^{\infty}(S) \nonumber\\
f \mapsto \partial_n \phi_f, \nonumber
\end{eqnarray}
where $\phi_f$ is the solution of the Dirichlet problem, i.e.\ it satisfies $\Delta
\phi_f(\vec{r}) =0$ inside and $\phi_f(\vec{r})=f(\vec{r})$ on the surface $S$ of the
particle. The Dirichlet-Neumann operator outside the particle ($D_+$) is defined
in a similar way.\\

 With the help of the operators $D_-$ and $D_+$ the problem in eqs.~(\ref{eq:poisson})-(\ref{eq:bc2})
can be recast in a compact manner as
\begin{equation}
\epsilon D_- f = D_+ f
\end{equation}
where $f$ is the restriction of $\phi_\pm$ to $S$.
Nontrivial solutions of this equation can be found for such $\epsilon$ for which
\begin{equation} \label{eq:ker reformulate}
\ker(\epsilon D_- - D_+) \neq 0 \quad \Leftrightarrow \quad \ker(\epsilon D_-
D_+^{-1}-\mathds{1}) \neq 0.
\end{equation}
Note that $D_+$ is an invertible operator. Thus, the desired values of $\epsilon$ are the inverse eigenvalues of
the operator $D_-D_+^{-1}$, i.e.,
\begin{equation}
\epsilon=\frac{1}{\text{eigenvalue}\{D_-D_+^{-1}\}} \label{eq:eigenv}.
\end{equation}

Hence, by means of the Dirichlet-Neumann operators we have reformulated the boundary
value problem for the determination of the surface modes as an eigenvalue
problem.

As a side remark, note that $D_-D_+^{-1}$ is not selfadjoint with respect to the
standard scalar product, although both $D_\pm$ are. This can be remedied either
by considering the non-standard scalar product $(f,g)=\int_S (D_+
f)(x)\overline{g(x)}\,dS(x)$ instead or by using $D_+^{-1/2} D_- D_+^{-1/2}$ in
(\ref{eq:ker reformulate}) instead of $D_-D_+^{-1}$, which is selfadjoint for
the standard scalar product and therefore shows the reality and completeness of
the spectrum. The discreteness of the spectrum of $D_-D_+^{-1}$ is less obvious and
will be shown below for a bounded surface $S$.
\section{Surface modes of a half-space}
\label{sec:half-space}
As an illustration of the method developed above we determine the surface modes
of a half-space. Since the resonances can be found at such $\epsilon$ which
are equal to the inverse eigenvalues of $D_-D_+^{-1}$ we first derive the
Dirichlet-Neumann operators $D_-$ and $D_+$ for a half-space. The surface $S$ of the
material, located at $z<0$ with the permittivity $\epsilon$, is chosen to be the
$x$-$y$-plane, so that we have to solve the Laplace equation
\begin{equation}
\label{eq:halfspace}
\Delta \phi_{\mp}(x,y,z)=0
\end{equation}
 for $z<0$ and $z>0$, resp., with the Dirichlet boundary condition
\begin{equation}
\label{eq:hs_bc}
\phi_\mp(x,y,0)=f(x,y);
\end{equation}
outside the material, for $z>0$, we assume vacuum. In order to solve
eqs.~(\ref{eq:halfspace}) and (\ref{eq:hs_bc}) we use the Fourier transform
in the $x$- and $y$-coordinates,
\begin{equation}
\hat{f}(\vec{\xi},z)=\int\!\! {\rm d} \vec{x} \, {\rm e}^{-{\rm i}\vec{x} \cdot
  \vec{\xi}}f(\vec{x},z),
\end{equation}
with the two Fourier variables $\vec{\xi}$ and $\vec{x}=(x,y)$. Fourier
transformation of eqs.~(\ref{eq:halfspace}) and (\ref{eq:hs_bc}) gives
\begin{equation}
\label{eq:ft}
(-|\vec{\xi}|^2+\partial_z^2)\hat{\phi}_\mp(\vec{\xi},z)=0 \quad \text{and} \quad
  \hat{\phi}_\mp(\vec{\xi},0)=\hat{f}(\vec{\xi}).
\end{equation}
From this one obtains the solution
\begin{equation}
\hat{\phi}_\mp(\vec{\xi},z)=e^{\pm|\vec{\xi}|z}\hat{f}(\vec{\xi}),
\end{equation}
(using the additional boundary condition that $\phi$ vanishes for
$|z| \rightarrow \infty$)
from which it becomes obvious that the surface modes are confined to a small neighbourhood of  the surface
of the material.\\

For the case considered the outward directional derivative is simply given by
$\partial_z$, so that the Dirichlet-Neumann operator $D_-$ can be stated as
\begin{equation}
\label{eq:D-}
D_-(f)=\frac{1}{(2 \pi)^2}\int \! \text{d}\vec{\xi} e^{i(\vec{x}\cdot\vec{\xi})}|\vec{\xi}|\hat{f}(\vec{\xi})
\end{equation}
and $D_+$ as
\begin{equation}
\label{eq:D+}
D_+(f)=-\frac{1}{(2 \pi)^2}\int \! \text{d}\vec{\xi} e^{i(\vec{x}\cdot\vec{\xi})}|\vec{\xi}|\hat{f}(\vec{\xi}).
\end{equation}
In particular, for a half-space we have $D_-=-D_+$ and therefore
$D_-D_+^{-1}=-\mathds{1}$. Since the only eigenvalue of the unit operator
$\mathds{1}$ is $1$ we obtain the well-known result that $\epsilon=-1$ for the surface modes of a half-space.
\section{High-order surface modes}
\label{sec:high}
Before we study the convergence of the eigenvalues of the operator $D_-
D_+^{-1}$ in rigoruos mathematical terms, we briefly discuss the physical
expectation. For example, considering a sphere surrounded by vacuum it is well
known that the resonances of the surface modes are given by \cite{Raether}
\begin{equation}\label{eq:sphere}
\epsilon_k=-\frac{k+1}{k}
\end{equation}
for $k =1,2,\dots$. Obviously, for $k=1$, as corresponding to the dipole mode
 of the sphere, the resonance can be found at $\epsilon=-2$. On the other hand,
 in the limiting case of $k \rightarrow \infty$, corresponding to the multipole
 mode of infinite order, the resonance occurs at $\epsilon=-1$, i.e., one
 retrieves the half-space result deduced above. This convergence property can be
 understood from the physicist's point of view as follows: The multipole modes
 of order $k \gg 1$ are related to fields which vary on length scales much
 smaller than the radius of the sphere; the higher the order $k$ the smaller the
 length scale. Therefore, these high-order modes cannot distinguish
 the sphere from a half-space, since the sphere can be considered as
 locally flat on the length scale of these modes. Hence, the convergence to
 $\epsilon=-1$ for $k \to \infty$ is not restricted to spheres, but should be expected for
 arbitrary particles. In the following we prove this property in strict
 mathematical terms.\\

To this end we first briefly introduce the concept of pseudo-differential
operators, for a more detailed introduction to this topic we refer the reader to
\cite{taylor}.  The definition of a general pseudo-differential operator $P$ with
{\it symbol} $p(\vec{x},\vec{\xi})$ operating on the function $u(\vec{x})$
can be stated as (with $\vec{x},\vec{\xi} \in \mathds{R}^n$)
\begin{equation}
[Pu](\vec{x})=\frac{1}{(2 \pi)^n} \int \!\!\!
\text{d}\vec{\xi} \, {\rm e}^{{\rm i}\vec{x}\cdot
\vec{\xi}}p(\vec{x},\vec{\xi})\hat{u}(\vec{\xi}).
\end{equation}
From this definition it can be seen that if the symbol $p(\vec{x},\vec{\xi})$ is
a polynomial in $\vec{\xi}$ the operator $P$ is a conventional differential
operator, which has constant coefficients if $p$ is independent of $\vec x$.
However, for pseudo-differential operators one admits more general symbols: The
smooth function $p$ is required to have an asymptotic expansion as
$\xi\to\infty$ of the form $p(\vec x, \vec \xi) \sim p_m (\vec x,\vec\xi) +
p_{m-1}(\vec x,\vec\xi) + p_{m-2}(\vec x, \vec \xi)+\cdots$, where each $p_{m-i}$
is positively homogeneous in $\vec\xi$ of degree $m-i$, i.e.\ satisfies
$p_{m-i}(\vec x,\vec\xi) = |\vec\xi|^{m-i}p_{m-i}(\vec x, \hat{\vec\xi})$ for
$\vec\xi\neq 0$, where $\hat{\vec\xi}=\vec\xi/|\vec\xi|$.\footnote{Sometimes one
considers even more general symbols, and then the ones we defined are called
classical symbols of type $(1,0)$, see \cite{taylor}.} The number $m$ can be an
arbitrary real number and is called the {\em order} of $P$. The leading term
$p_m$ is called the {\em principal symbol} of $P$ and denoted as $\sigma_m(P)$. In
the case of a differential operator with $p(\vec
x,\vec{\xi})=\sum\limits_{|\beta| \leq m} \alpha_{\beta}(\vec x)
\vec{\xi}^{\beta}$ (where $\beta\in\mathbb{N}_0^n$ is a multi-index), the
principal symbol is $\sigma_m(P)=\sum\limits_{|\beta|=m}\alpha_{\beta}(\vec x)
\vec{\xi}^{\beta}$.\\

We defined pseudo-differential operators as acting on functions of $\vec
x\in\mathds{R}^n$, which seems essential since the Fourier transform was used in
the definition. However, by means of local coordinates one may define
pseudo-differential operators on a manifold (for example, the surface $S$), and
then the principal symbol is independent of the choice of coordinates if
$\vec\xi$ is interpreted as a covector.

A basic (albeit non-obvious) property of pseudo-differential operators is that
if $P$ and $Q$ are pseudo-differential operators of orders $m,l$, respectively,
then their composition $PQ$ is a pseudo-differential operator of order $m+l$,
with principal symbol $\sigma_{m+l}(PQ) = \sigma_m(P)\sigma_l(Q)$. The identity
operator has order $0$ with principal symbol $\sigma_0(\mathds{1})=1$, therefore
if $P$ is invertible then the principal symbol of its inverse is
$\sigma_{-m}(P^{-1}) = (\sigma_m(P))^{-1}$.

In order to prove the shape-independent convergence of the high order surface
modes we need one more well-known result (see \cite{taylor}): The
Dirichlet-Neumann operators $D_\pm$ are pseudo-differential operators on $S$
whose principal symbols are the same as in the case of a
half-space~(\ref{eq:D-}) and~(\ref{eq:D+}):
\begin{equation}
\sigma_1(D_-)=|\vec{\xi}|,
\qquad
\sigma_1(D_+)=-|\vec{\xi}|.
\end{equation}
From this we get
$\sigma_{-1}(D_+^{-1})=-\frac{1}{|\vec{\xi}|} $
and then
\begin{equation}
\sigma_{0}(D_-D_+^{-1})=-1.
\end{equation}
This implies that the zeroth order operator $R:=D_- D_+^{-1} + \mathds{1}$ has
the principal symbol $\sigma_0(R)=-1+1=0$, hence is in fact a
pseudo-differential operator of order $-1$. Now any pseudo-differential operator
of negative order on a bounded surface is a compact operator, and the spectral
theory of compact operators implies that the eigenvalues $r_k$ of $R$ form a a
sequence converging to zero as $k\to\infty$. The eigenvalues of $D_- D_+^{-1}= -
\mathds{1}+R$ are $-1+r_k$, so the resonances obey
\begin{equation}
\epsilon_k=(-1+r_k)^{-1} \to -1\quad\text{ as }k\to\infty.
\end{equation}

We emphasise that this limit is independent of the shape of the particle and
equal to the result for the half space. The physical expectation based on our
consideration of curvature becoming invisible to leading order at small scales
is reflected in this argument by the fact that the principal symbol of the
Dirichlet-Neumann operator is independent of the shape.

We remark that with more refined methods of spectral asymptotics for
pseudo-differential operators a more precise asymptotic result can be obtained
which explains the correction $-\frac1k$ in the formula $\epsilon_k = -1
-\frac1k$ for the sphere, (\ref{eq:sphere}).

Finally, we mention a different reduction of the plasmonic eigenvalue problem
(\ref{eq:poisson})-(\ref{eq:bc2}) to an eigenvalue problem on the surface $S$,
derived by the method of layer potentials in \cite{Abajo_Aizpurua},
\cite{Abajo_Aizpurua2}: For $\vecr,\vecr'$ on $S$ let
$$ F(\vecr,\vecr') = -\frac1{2\pi} \frac{n(\vecr) \cdot
(\vecr-\vecr')}{|\vecr-\vecr'|^3}$$ 
where $n(\vecr)$ is the outer unit
normal. Then $\epsilon=\frac{\lambda-1}{\lambda+1}$ for the eigenvalues
$\lambda$ in the equation
$$ \int_S F(\vecr,\vecr')\, \sigma(\vecr')\, d^2\vecr' = \lambda \sigma (\vecr),
\quad \vecr \in S.$$ 
Here $\sigma$ is the outward normal derivative
$\partial_n\phi$ (which is proportional to the induced surface charge
density). The operator on the left is compact (in fact, a pseudodifferential
operator of order $-1$), hence $\lambda_k\to 0$, and this shows $\epsilon_k\to
-1$ again. For a proof of this fact see \cite{taylor} where also the relation of
this operator and the Dirichlet-Neumann operators is discussed.

\section{Conclusion}
In this paper, we have applied methods of microlocal analysis to the study
of plasmon resonances of arbitrarily shaped nanoparticles. We have first
reformulated the boundary value problem for such resonances as an eigenvalue
problem on the particle's surface. The fact that the Dirichlet-Neumann
operators, which occur naturally in this context, are pseudo-differential
operators then allows one to take advantage of the rich amount of knowledge
available for these objects, and thereby to analyse the properties of the
surface modes in rigorous mathematical terms. The remarkable ease with which
the well-known result $\epsilon = -1$ for the surface modes of a half-space
has been recovered here is a clear indication for the power of this approach.
Moreover, we have used the eigenvalue equation for proving that the limit
of the high order modes is independent of the shape of the particle. As
expected on the grounds of intuitive physical arguments, the high-order
modes converge locally to the half-space modes. While this result itself
appears natural, what matters here is the mathematical toolbox by which
it has been obtained, which may be still somewhat unfamiliar to physicists,
but which offers great conceptual clarity and flexibility. Thus, we hope that
the approach suggested in this paper will prove useful for obtaining further
insight into mathematical problems ocurring in plasmonics.

\ack
This work was supported in part by the DFG through Grant. No. KI 438/8-1.


\end{document}